\documentclass[aps,pra,showpacs,superscriptaddress,reprint,showkeys]{revtex4-1}
\usepackage{amsfonts}
\usepackage{graphicx}
\usepackage{latexsym}
\usepackage{makeidx}
\usepackage{amsmath}
\usepackage{amssymb}
\usepackage{graphicx}
\usepackage{soul}
\usepackage{color}

\begin{document}

\title{Effective Landau-Zener transitions in circuit dynamical Casimir
effect with time-varying modulation frequency}
\date{\today }
\author{A. V. Dodonov}
\affiliation{Institute of Physics and International Center for Condensed
Matter Physics, University of Brasilia, 70910-900, Brasilia, Federal
District, Brazil}
\affiliation{Dipartimento di Fisica e Chimica,
Universit\`a degli Studi di Palermo, Via Archirafi 36, I-90123 Palermo,
Italy}
\author{B. Militello}
\affiliation{Dipartimento di Fisica e Chimica, Universit\`a degli Studi di
Palermo, Via Archirafi 36, I-90123 Palermo, Italy}
\affiliation{I.N.F.N., Sezione di Catania}
\author{A. Napoli}
\affiliation{Dipartimento di Fisica e Chimica, Universit\`a degli Studi di
Palermo, Via Archirafi 36, I-90123 Palermo, Italy}
\affiliation{I.N.F.N., Sezione di Catania}
\author{A. Messina}
\affiliation{Dipartimento di Fisica e Chimica, Universit\`a degli Studi di
Palermo, Via Archirafi 36, I-90123 Palermo, Italy}
\affiliation{I.N.F.N., Sezione di Catania}

\begin{abstract}
We consider the dissipative single-qubit circuit QED architecture in which
the atomic transition frequency undergoes a weak external time-modulation.
For sinusoidal modulation with linearly varying frequency we derive
effective Hamiltonians that resemble the Landau-Zener problem of finite
duration associated to a two- or multi-level systems. The corresponding
off-diagonal coupling coefficients originate either from the rotating or the
counter-rotating terms in the Rabi Hamiltonian, depending on the values of
the modulation frequency. It is demonstrated that in the dissipationless
case one can accomplish almost complete transitions between the eigenstates
of the bare Rabi Hamiltonian even for relatively short duration of the
frequency sweep. To assess the experimental feasibility of our scheme we
solved numerically the phenomenological and the microscopic quantum
master equations in the Markovian regime at zero temperature. Both models
exhibit qualitatively similar behavior and indicate that photon generation
from vacuum via effective Landau-Zener transitions could be implemented with
the current technology on the timescales of a few microseconds. Moreover,
unlike the harmonic dynamical Casimir effect implementations, our proposal does not require the
precise knowledge of the resonant modulation frequency to accomplish
meaningful photon generation.
\end{abstract}

\maketitle

\section{Introduction}

The area of circuit Quantum Electrodynamics (circuit QED) offers
unprecedented possibilities to manipulate \emph{in situ} the properties of
mesoscopic systems composed of superconducting artificial atoms interacting
with the Electromagnetic field inside the waveguide resonator on a chip \cite{ref:Wallraff2004, scho,ref:Nori2011}.
Along with practical application for Quantum Information Processing (QIP) this
solid-state architecture also allows for the experimental study of some of
the most fundamental physical processes, such as the light--matter
interaction at the level of a few photons. Due to the relative ease to
achieve the {\em strong coupling regime}, in which the coherent light--matter
coupling rate is much larger than the system damping and dephasing rates \cite{ref:Wallraff2004,resolve},
this setup can probe novel phenomena with tiny coupling rates. One example
is the implementation of the {\em dynamical Casimir effect} (DCE) and associated
phenomena using actively controlled artificial atoms, which may serve both
as the source and the detector of modulation-induced radiation \cite{jpcs,roberto,PLA,ser1,ser2,ser3,ser4}. We recall that DCE is a
common name ascribed to the processes in which photons are generated from
vacuum due to the external time-variation of boundary conditions for some
field \cite{book,vdodonov,revDal,nori}. For the usual Electromagnetic case this corresponds to the fast
motion of a mirror or modulation of the dielectric properties of the mirror
/ intracavity medium \cite{SV1}. Analogs of DCE were recently verified experimentally
in the setups resembling a single-mirror \cite{DCE1} and a lossy cavity \cite{DCE2},
where the modulation of the boundary conditions was achieved by
threading a time-dependent magnetic flux through SQUIDs (superconductive
quantum interference devices) located inside the coplanar waveguide. These experiments stimulated new
theoretical research on role of dynamical Casimir physics in quantum-information processing, quantum simulations and engineering of nonclassical states of light and matter \cite{relativistic,processing,nonclassical,entangles,arrays,daniel}.

Recent studies have indicated that DCE could be implemented even using a
single two-level atom (qubit) with time-dependent parameters, such as the
transition frequency or the atom--field coupling strength \cite{jpcs,libe,JPA,igor,diego}. Generation of
excitation from vacuum occurs due to the counter-rotating terms in the Rabi
Hamiltonian, which for many years had been neglected under the Rotating Wave
Approximation (RWA). Moreover, single-atom DCE carries some new
characteristics, such as the atom--field entanglement, saturation in the
number of created photons (due to intrinsic nonlinearities associated with
non-harmonic spectrum of the composite system) and emergence of additional
resonant modulation frequencies \cite{igor}. Since the ultra-strong coupling regime \cite{usc1,usc2,usc3,usc4} (for
which the atom--field coupling rate is comparable to the cavity and atomic
frequencies) is experimentally demanding, especially in nonstationary
configurations, here we assume moderate values of the coupling strength, which in
turn imply small transition rates for the phenomena originating from the
counter-rotating terms. Hence the dissipation must be sufficiently weak and
the frequency of modulation must be tuned with a high precision,
typically of the order of $10\,-100$ kHz for modulation frequencies in the
GHz range, that ultimately should be determined experimentally or
numerically. For the harmonic modulation of system parameters the observable
quantities, e. g., the average photon number or the atomic excitation probability,
exhibit a oscillating behavior as function of time \cite{jpcs,diego}. Therefore the duration of
modulation must also be minutely adjusted to grasp a meaningful amount of
excitations, since the detection is typically carried out when the
modulation has ceased.

In this paper we propose a simple scheme to implement the
phenomena induced by the counter-rotating terms that does not require an accurate knowledge
of the resonance frequencies and is little
sensitive to the duration of the perturbation. Our method is based
on the {\em adiabatic} variation of the modulation frequency of the atomic
level splitting through the expected resonance, when one can
steadily create excitations from vacuum and other initial states
without the aforementioned sporadic comebacks of the system to the
initial state. This phenomenon can be understood
in terms of  some effective two- or multi-level Hamiltonians with
constant nondiagonal couplings, dependent on the modulation depths,
and time-dependent diagonal terms, which are responsible for
{\em Landau-Zener processes}.

By the way, Landau-Zener transitions \cite{ref:Landau,ref:Zener,ref:Majorana,ref:Stuck} are a very fundamental phenomenon, since they occur every time a two-level system is subjected to a time-dependent Hamiltonian whose bare energies change with time and cross at some instant of time. In its original form, LZ model is characterized by bare energies changing linearly in time. Over the years, the original scheme has been extended in several directions: non linear time-dependence of the diagonal matrix elements of the Hamiltonian \cite{ref:nonlinear,wilhelm}, multilevel systems, even with n-fold crossings \cite{ref:multilevel1,ref:multilevel2}, and, of course, including environmental effects inducing dissipation and decoherence \cite{ref:CircQED3,ref:dissipation1,ref:dissipation2,ref:dissipation3}.
This ubiquitous model has been applied to describe similar dynamical situations in several physical scenarios. For example, it has been useful in such systems as  spinorial Bose-Einstein condensates \cite{ref:Condensates}, Josephson junctions \cite{ref:Joseph1,ref:Joseph2},  in optical lattices with cold atoms \cite{ref:lattices}, and even --- in its modified version know as \lq Hidden Crossing model\rq\, \cite{ref:Hidden} --- in classical optics \cite{ref:Bowmeester}. Also in the context of
Circuit QED, Landau-Zener tunneling has been extensively used to
suitably manipulate the state of a single qubit
\cite{ref:CircQED3,ref:CircQED1,ref:CircQED2}. In our case, Landau-Zener processes naturally
rise from the fact that the frequency sweep allows realizing a
series of resonances between the (time-varying frequency) external
perturbation and a series of couples of the Rabi-Hamiltonian eigenstates.

One possible downside of our method is the relatively long implementation times.
Nonetheless, we demonstrate that it can work with the current dissipative
parameters provided the modulation depth of the atomic transition frequency
is of the order of a few percent of its bare value. Besides, we show that the states generated from vacuum can be quite different from the squeezed vacuum state (SVS) that is typically generated for strictly harmonic modulations \cite{SV1,SV2}, so our scheme may find new applications in QIP.
%{\bf The analytical study of Landau-Zener transitions in the presence of decoherence and dissipation sources is a tough job. In this case, a numerical analysis is indeed preferable. }

This paper is organized as follows. In section \ref{sec1} we deduce the
effective Hamiltonians similar to the Landau-Zener physics in the resonant
and dispersive regimes of atom--field interaction. In section \ref{sec2} we
describe our approach to realistically account for the dissipation, with
thorough numerical analysis carried out in section \ref{sec3}. Finally, in
section \ref{sec4} we discuss the obtained results and present our
conclusions.

\section{Effective Landau-Zener sweeps}

\label{sec1}

We consider a single qubit in nonstationary circuit QED. Our starting point
is the Hamiltonian (we set $\hbar =1$)%
\begin{equation}
\hat{H}_{R}(t)=\omega _{0}\hat{n}+\frac{\Omega (t)}{2}\hat{\sigma}%
_{z}+g_{0}\left( \hat{a}+\hat{a}^{\dagger }\right) \left( \hat{\sigma}_{+}+%
\hat{\sigma}_{-}\right) ,  \label{H1}
\end{equation}%
where $\hat{a}$ and $\hat{a}^{\dagger }$ are cavity annihilation and
creation operators and $\hat{n}=\hat{a}^{\dagger }\hat{a}$ is the photon
number operator; $\hat{\sigma}_{+}=|e\rangle \langle g|$, $\hat{\sigma}%
_{-}=|g\rangle \langle e|$ and $\hat{\sigma}_{z}=|e\rangle \langle
e|-|g\rangle \langle g|$ are the atomic ladder operators, where $|g\rangle $
($|e\rangle $) denotes the atomic ground (excited) state. $\omega _{0}$ is
the cavity frequency, $\Omega $ is the atomic transition frequency and $g_{0}
$ is the atom--field coupling strength. For the sake of simplicity here we
consider the external modulation of the atomic transition frequency as $%
\Omega (t)=\Omega _{0}+\varepsilon _{\Omega }\sin [\eta (t)t+\phi _{\Omega }]
$, although it was shown that for $g_{0}\ll \omega _{0},\Omega _{0}$ the
weak modulation of any system parameter produces similar results. In this
work we suppose that the modulation frequency $\eta (t)$ may also slowly
change as function of time. It will be shown that in specific regimes we
obtain the two-level or multi-level effective Landau-Zener physics, so
initially unpopulated system states can be excited without undergoing
oscillations back to the initial state.

For $\varepsilon _{\Omega }\ll \Omega _{0}$ and $\varepsilon _{\Omega
}\lesssim g_{0}$ we can treat the external modulation as a perturbation that
drives the transitions between the bare eigenstates of the Rabi Hamiltonian,
$\hat{H}_{R}^{(0)}|R_{i}\rangle =E_{i}|R_{i}\rangle $, where%
\begin{equation}
\hat{H}_{R}^{(0)}=\omega _{0}\hat{n}+\frac{\Omega _{0}}{2}\hat{\sigma}%
_{z}+g_{0}\left( \hat{a}+\hat{a}^{\dagger }\right) \left( \hat{\sigma}_{+}+%
\hat{\sigma}_{-}\right)  \label{H2}
\end{equation}%
and $E_{i}$ increases with the index $i$. Expanding the wavefunction
corresponding to the Hamiltonian (\ref{H1}) as%
\begin{equation}
|\psi (t)\rangle =\sum_{i}A_{i}(t)e^{-itE_{i}}|R_{i}\rangle
\end{equation}%
the probability amplitudes obey the coupled differential equations%
\begin{equation}
i\dot{A}_{j}(t)=\frac{\varepsilon _{\Omega }\sin \left( \eta t+\phi _{\Omega
}\right) }{2}\sum_{i}A_{i}(t)e^{-it\left( E_{i}-E_{j}\right) }\langle R_{j}|%
\hat{\sigma}_{z}|R_{i}\rangle .  \label{kiko}
\end{equation}%
Thus one can induce the coherent coupling between the Rabi dressed states $%
\{|R_{i}\rangle ,|R_{j}\rangle \}$ by setting the modulation frequency
roughly equal to $|E_{i}-E_{j}|$. The remaining rapidly oscillating terms
can be neglected according to the RWA
approach, which sets the criteria for the validity of the resulting
effective equations. In addition, the RWA method introduces small intrinsic
frequency shifts in the final equations, which slightly alter the resonant
modulation frequency \cite{JPA,igor}. One of the advantages of the present scheme
is that such frequency shifts are completely irrelevant for
the experimental implementation.

For the weak qubit--field coupling considered here, $g_{0}\ll \omega _{0}$, we can find the
approximate spectrum of the Rabi Hamiltonian by performing the
unitary
transformation \cite{blais}%
\begin{equation}
\hat{U}_{R}=\exp \left[ \Lambda \left( \hat{a}\hat{\sigma}_{-}-\hat{a}%
^{\dagger }\hat{\sigma}_{+}\right) +\xi \left( \hat{a}^{2}-\hat{a}^{\dagger
2}\right) \hat{\sigma}_{z}\right] ,
\end{equation}%
where $\Lambda \equiv g_{0}/\Delta _{+}$, $\xi \equiv \Lambda g_{0}/2\omega
_{0}$ and $\Delta _{\pm }=\omega _{0}\pm \Omega _{0}$. To the first order in
$\Lambda $ we get the Bloch-Siegert Hamiltonian~\footnote{The Hamiltonian (\ref{lula}) may be used even in the ultra-strong coupling regime, provided
$g_0$ is small with respect to $\Delta_+$.}%
\begin{eqnarray}
\hat{H}_{BS} &=&\hat{U}_{R}^{\dagger }\hat{H}_{R}^{(0)}\hat{U}_{R}=\left(
\omega _{0}+\delta _{+}\hat{\sigma}_{z}\right) \hat{n}  \notag \\
&&+\frac{\Omega _{0}+\delta _{+}}{2}\hat{\sigma}_{z}+g_{0}\left( \hat{a}\hat{%
\sigma}_{+}+\hat{a}^{\dagger }\hat{\sigma}_{-}\right) \label{lula}.
\end{eqnarray}%
Hence the approximate eigenvalues of $\hat{H}_{R}^{(0)}$ are
\begin{equation}
E_{0}=-\left( \Omega _{0}+\delta _{+}\right) /2  \label{v1}
\end{equation}%
\qquad\
\begin{equation}
E_{n>0,\pm }=\omega _{0}n-\frac{\omega _{0}+\delta _{+}}{2}\pm \frac{1}{2}%
\sqrt{ [\tilde{\Delta} _{-}(n)] ^{2}+4g_{0}^{2}n}\,,
\end{equation}%
where $\delta _{\pm }\equiv g_{0}^{2}/\Delta _{\pm }$, $\tilde{\Delta}_{-}(n)\equiv \Delta _{-}-2\delta _{+}n$ and the integer $n$ is
the number of total excitations associated to $\hat{H}_{BS}$. The
approximate eigenstates of (\ref{H2}) are $|R_{i}\rangle =\hat{U}%
_{R}|\Upsilon _{i}\rangle $, where $|\Upsilon _{i}\rangle $ are the
eigenstates of $\hat{H}_{BS}$:
\begin{equation}
|\Upsilon _{0}\rangle =|g,0\rangle\,,
\end{equation}%
\begin{equation}
|\Upsilon _{n>0,+}\rangle =\sin \theta _{n}|g,n\rangle +\cos \theta
_{n}|e,n-1\rangle\,,
\end{equation}%
\begin{equation}
|\Upsilon _{n>0,-}\rangle =\cos \theta _{n}|g,n\rangle -\sin \theta
_{n}|e,n-1\rangle\,,
\end{equation}%
%\begin{equation}
%\theta _{n>0}=\arctan \left( \frac{\Delta _{-}-2\delta _{+}n+\sqrt{\left(
%\Delta _{-}-2\delta _{+}n\right) ^{2}+4g_{0}^{2}n}}{2g_{0}\sqrt{n}}\right) .
%\label{v3}
%\end{equation}
\begin{equation}
\theta _{n>0}=\arctan \left( \frac{\tilde{\Delta}_{-}(n)+\sqrt{[\tilde{\Delta}_{-}(n)]^{2}+4g_{0}^{2}n}}{2g_{0}\sqrt{n}}\right)\,.
\label{v3}
\end{equation}

Contrary to the standard studies on DCE and related resonant effects, where
the precise knowledge of spectrum is necessary to accomplish the desired
transitions \cite{libe,vedral,igor,diego}, in this work only an approximate
knowledge of the spectrum is enough to achieve the phenomena of interest.
Therefore the lowest order eigenvalues and eigenstates derived above are
sufficient for our purposes. The essence of our proposal is most clearly
seen in specific regimes of the system parameters, as illustrated below for
the resonant and dispersive regimes.

\subsection{Resonant regime}

First we consider the {\em resonant regime}, $\Delta _{-}=0$, and assume a small
number of excitations, $\Lambda ^{2}n\ll 1$. For the initial system ground
state $|R_{0}\rangle $ and the modulation frequency
\begin{equation}
\eta =2\omega _{0}\pm g_{0}\sqrt{2}-\nu \left( t\right)  \label{she}
\end{equation}%
one can show [by neglecting the rapidly oscillating terms in equation (\ref%
{kiko})] that to the lowest order in $\Lambda $ the dynamics is described by
the effective Hamiltonian%
\begin{eqnarray}
\hat{H}_{i} &=&E_{0}|R_{0}\rangle \langle R_{0}|+E_{2,\pm }|R_{2,\pm
}\rangle \langle R_{2,\pm }| \\
&&\pm \left( ig_{0}\frac{\sqrt{2}}{4}\frac{\varepsilon _{\Omega }^{\prime }}{%
\Delta _{+}}e^{it\left( E_{2,\pm }-E_{0}-\nu \left( t\right) \right)
}|R_{0}\rangle \langle R_{2,\pm }|+h.c.\right) .  \notag
\end{eqnarray}%
Here we defined the complex modulation depth $\varepsilon _{\Omega }^{\prime
}\equiv \varepsilon _{\Omega }e^{i\phi _{\Omega }}$ and $\nu \left( t\right)
$ is a small time-dependent function, $|\nu \left( t\right) |\ll g_{0}$, to
be specified later. This Hamiltonian can also be obtained after cumbersome calculations using the method of
\cite{JPA,igor}, though in this paper it is
derived just in a few lines~\footnote{Although the present method is more practical to deduce the effective Hamiltonian, the approach of \cite{JPA} is more suitable to account for nonlinear effects and to estimate the intrinsic frequency shifts. Such shifts are irrelevant when $\nu (t)$ undergoes a linear sweep.}. Performing the time-dependent unitary transformation%
\begin{eqnarray}
\hat{S}\left( t\right) &=&\exp \left\{ -it\left[ \left( E_{0}+\frac{\nu
\left( t\right) }{2}\right) |R_{0}\rangle \langle R_{0}|\right. \right.
\notag \\
&&\left. \left. +\left( E_{2,\pm }-\frac{\nu \left( t\right) }{2}\right)
|R_{2,\pm }\rangle \langle R_{2,\pm }|\right] \right\}
\end{eqnarray}%
we obtain the \lq interaction-picture\rq\
effective Hamiltonian%
\begin{eqnarray}
\hat{H}_{f} &\equiv &-i\hat{S}^{\dagger }\frac{d\hat{S}}{dt}+\hat{S}%
^{\dagger }\hat{H}_{i}\hat{S}  \label{vortex} \\
&=&\frac{V(t)}{2}\left( |R_{2,\pm }\rangle \langle R_{2,\pm }|-|R_{0}\rangle
\langle R_{0}|\right) \pm \left( i\beta |R_{0}\rangle \langle R_{2,\pm
}|+h.c.\right) ~  \notag
\end{eqnarray}%
\begin{equation}
V(t)\equiv \nu \left( t\right) +t\dot{\nu}\left( t\right) ~,~\beta =\frac{1}{%
\sqrt{2}}g_{0}\frac{\varepsilon _{\Omega }^{\prime }}{2\Delta _{+}}.
\end{equation}%
Notice that this Hamiltonian only holds for the modulation frequency (\ref%
{she}), and $\beta $ is nonzero due to the presence of the counter-rotating
terms in the Rabi Hamiltonian (\ref{H2}).

When $\nu \left( t\right) $ is a linear function of time the equation (\ref%
{vortex}) describes the standard two-level Landau-Zener problem, so for
adiabatic variation of $V(t)$ across the avoided-crossing one can transfer
steadily the population from the initial state $|R_{0}\rangle $ to the final
one $|R_{2,\pm }\rangle $. For the linear variation of $V(t)$ from $-\infty $
to $+\infty $ the probability of such transition is $1-\exp \left( -\pi
\beta ^{2}/|\dot{\nu}|\right) $, which is close to one provided we have $|%
\dot{\nu}|\ll \pi \beta ^{2}$. However, the equation (\ref{vortex}) is only
valid for small values of $|\nu |$, so our scheme corresponds to the finite
duration Landau-Zener sweeps in which we have $|\nu (t)|\leq K|\beta |$,
where $K$ is of the order of $10$. Nonetheless, as shown later, in this way
we can achieve the desired transition with sufficiently high probability and
compensate for the ignorance in the knowledge of the exact eigenvalues of
the system Hamiltonian.

\subsection{Dispersive regime}

The {\em dispersive regime} is defined as $g_{0}\sqrt{n}\ll \left\vert \Delta
_{-}\right\vert /2$ for all relevant values of $n$, and we also assume the
standard condition $|\Delta _{-}|\ll \omega _{0}$. Repeating the above
reasoning one finds that for the initial ground state and the modulation
frequency%
\begin{equation}
\eta =\Delta _{+}-2\left( \delta _{-}-\delta _{+}\right) +4\alpha -\nu
(t)~,~\alpha \equiv \frac{g_{0}^{4}}{\Delta _{-}^{3}}
\end{equation}%
the interaction-picture effective Hamiltonian reads%
\begin{eqnarray}
\hat{H}_{f} &=&\frac{V(t)}{2}\left( |R_{2,-\mathcal{D}}\rangle \langle R_{2,-%
\mathcal{D}}|-|R_{0}\rangle \langle R_{0}|\right)  \notag \\
&&-\mathcal{D}\left( i\beta |R_{0}\rangle \langle R_{2,-\mathcal{D}%
}|+h.c.\right) ~
\end{eqnarray}%
\begin{equation}
\beta =g_{0}\frac{\varepsilon _{\Omega }^{\prime }}{2\Delta _{+}}~,
\end{equation}%
where $\mathcal{D}=\pm$, being the sign of $\Delta _{-}/\left\vert \Delta _{-}\right\vert$.
Thus one can achieve the steady population transfer from $|R_{0}\rangle $
to $|R_{2,-\mathcal{D}}\rangle $, which corresponds approximately to the
transition $|g,0\rangle \rightarrow |e,1\rangle $. For $\nu (t)=0$ this
behavior was previously named Anti-Jaynes-Cummings regime \cite{jpcs,roberto}
or the blue-sideband transition \cite{blais}.

On the other hand, for
\begin{equation}
\eta =2\omega _{0}+2\left( \delta _{-}-\delta _{+}\right) -4\alpha -\nu (t)
\end{equation}%
we obtain an analog of the DCE Hamiltonian in the presence of the Kerr
nonlinearity \cite{igor}%
\begin{eqnarray}
\hat{H}_{f} &=&\sum_{n=0}^{n_{\max }}\left[ \left( \frac{V-2\alpha \left(
n-2\right) }{2}\right) n|R_{n,\mathcal{D}}\rangle \langle R_{n,\mathcal{D}%
}|\right.  \label{mart} \\
&&\left. +\left( i\sqrt{\frac{\left( n+1\right) \left( n+2\right) }{2}}\beta
|R_{n,\mathcal{D}}\rangle \langle R_{n+2,\mathcal{D}}|+h.c.\right) \right] ~
\notag
\end{eqnarray}%
\begin{equation}
\beta =\delta _{-}\frac{\varepsilon _{\Omega }^{\prime }}{\sqrt{2}\Delta _{+}%
}.
\end{equation}%
Here we defined $|R_{0,\mathcal{D}}\rangle \equiv |R_{0}\rangle $, and $%
n_{\max }$ denotes the limiting value for the validity of the dispersive
regime. If $|\beta |\gtrsim |\alpha |$ we get the \lq multi-level\rq\ Landau-Zener
problem, in which one can couple several dressed states $|R_{2n,\mathcal{D}%
}\rangle \approx |g,2n\rangle $ at each avoided-crossing (although the avoided crossings for different pairs of levels do not coincide exactly), so one can
asymptotically create several photons from the initial vacuum state $%
|g,0\rangle $.

Our approach is not limited to the initial ground state. For example, for
the initial state with a finite number of excitations one can apply the
low modulation frequency%
\begin{equation}
\eta =\left\vert \Delta _{-}-2\delta _{+}m\right\vert +2\left\vert \delta
_{-}\right\vert m-2\left\vert \alpha \right\vert m^{2}-\nu (t)~.
\end{equation}%
For $\varepsilon _{\Omega }\lesssim \Delta _{-}$ we find the
interaction-picture effective Hamiltonian%
\begin{eqnarray}
\hat{H}_{f} &=&\sum_{n=1}^{n_{\max }}\left[ \left( \frac{\mathcal{D}%
V-2\left( \delta _{-}-\delta _{+}\right) \left( m-n\right) +2\alpha \left(
m^{2}-n^{2}\right) }{2}\right) \right.   \notag \\
&&\times \left( |R_{n,\mathcal{D}}\rangle \langle R_{n,\mathcal{D}}|-|R_{n,-%
\mathcal{D}}\rangle \langle R_{n,-\mathcal{D}}|\right)   \notag \\
&&\left. +\left( i\frac{\sqrt{n}}{2}\beta |R_{n,-\mathcal{D}}\rangle \langle
R_{n,\mathcal{D}}|+h.c.\right) \right] ~  \label{domi}
\end{eqnarray}%
\begin{equation}
\beta =g_{0}\frac{\varepsilon _{\Omega }^{\prime \left( \mathcal{D}\right) }%
}{\Delta _{-}}~,
\end{equation}%
where $\varepsilon ^{\prime (+)}\equiv \varepsilon ^{\prime }$ and $%
\varepsilon ^{\prime }{}^{(-)}\equiv \varepsilon ^{\prime \ast }$. The nondiagonal terms in the
Hamiltonian (\ref{domi}) rely only on the rotating terms in the Rabi
Hamiltonian, so the parameter $\Lambda $ does not appear in the
coupling coefficient $\beta $. For $\varepsilon _{\Omega }\sqrt{m}\ll g_{0}$
and $|\nu |_{\max }\lesssim 10|\beta |$ we achieve the coupling only between
the states $\{|R_{m,\mathcal{D}}\rangle ,|R_{m,-\mathcal{D}}\rangle \}$,
that corresponds approximately to the steady transition $|g,m\rangle
\rightarrow |e,m-1\rangle $. For larger variations of $|\nu |$ several
dressed states may become successively coupled during the frequency sweep.

\section{Account of dissipation}

\label{sec2}

For open quantum systems the dynamics must be described by the master
equation (ME) for the system density operator%
\begin{equation}
d\hat{\rho}/dt=-i[\hat{H}_{R}(t),\hat{\rho}]+\mathcal{\hat{L}}\hat{\rho},
\label{mer}
\end{equation}%
where $\mathcal{\hat{L}}$ is the Liouvillian superoperator whose form
depends on the details of system-reservoir interaction. In this work we do
not aim to develop a microscopic \emph{ab initio} model for dissipation in
nonstationary systems, instead we assess whether the effective Landau-Zener
transitions discussed in the previous section could be implemented in real
circuit QED architectures. So we use the simplest consistent dissipation
approaches available in the literature to evaluate numerically the dynamics
during the timescales of interest.

We consider independent reservoirs
for different processes, such as dissipation and pure
dephasing. Moreover, we assume that their correlation times are
much shorter than the system relevant timescales,
virtually zero, so
that we can treat the noise in the Markovian limit. Applying the
Davies-Spohn theory
\cite{ref:Davies1978}, we can consider that in a given time window
the bath sees the system as if it was governed by a
time-independent Hamiltonian, and if such time window is larger
than the typical correlation time of the bath, then the bath acts
on the system as if it was governed by a time-independent
Hamiltonian, time interval after time interval. If the transition
frequencies of the system are all different for any pair of eigenstates of $%
\hat{H}_{R}(t)$, which is true for the examples discussed below, then at
zero temperature the master equation reads \cite{blais}%
\begin{eqnarray}
\mathcal{\hat{L}}_{R}\bullet &=&\mathcal{D}\left[ \sum_{l}\Phi ^{l}|l\rangle
\langle l|\right] \bullet +\sum_{l,k\neq l}\Gamma _{\phi }^{lk}\mathcal{D}%
\left[ |l\rangle \langle k|\right] \bullet  \notag \\
&&+\sum_{l,k>l}\left( \Gamma _{\kappa }^{lk}+\Gamma _{\gamma }^{lk}\right)
\mathcal{D}\left[ |l\rangle \langle k|\right] \bullet \,,  \label{dave}
\end{eqnarray}%
where $\mathcal{D}[\hat{O}]\hat{\rho}\equiv \frac{1}{2}(2\hat{O}\hat{\rho}%
\hat{O}^{\dagger }-\hat{O}^{\dagger }\hat{O}\hat{\rho}-\hat{\rho}\hat{O}%
^{\dagger }\hat{O})$ is the Lindbladian superoperator and we use the
shorthand notation $|l\rangle $ to denote the time-dependent eigenstates of $%
\hat{H}_{R}(t)$, where the index $l$ increases with the eigenenergy $\lambda
_{l}(t)$. The time-dependent parameters of equation (\ref{dave}) are defined
as $\Phi ^{l}=[\gamma _{\phi }(0)/2]^{1/2}\sigma _{z}^{ll}$,$~~\Gamma _{\phi
}^{lk}=\gamma _{\phi }(\Delta _{kl})|\sigma _{z}^{lk}|^{2}/2$, $\Gamma
_{\kappa }^{lk}=\kappa (\Delta _{kl})|a^{lk}|^{2}$ and$~\Gamma _{\gamma
}^{lk}=\gamma (\Delta _{kl})|\sigma _{x}^{lk}|^{2}$. Here $\kappa (\varpi )$%
, $\gamma (\varpi )$ and $\gamma _{\phi }(\varpi )$ are the dissipation
rates corresponding to the resonator and qubit dampings and dephasing noise
spectral densities at frequency $\varpi $; we also defined the
time-dependent quantities $\Delta _{kl}\equiv \lambda _{k}(t)-\lambda
_{l}(t) $, $\sigma _{z}^{lk}=\langle l|\hat{\sigma}_{z}|k\rangle $, $%
a^{lk}=\langle l|(\hat{a}+\hat{a}^{\dagger })|k\rangle $ and $\sigma
_{x}^{lk}=\langle l|(\hat{\sigma}_{+}+\hat{\sigma}_{-})|k\rangle $.

From the lowest order approximate expressions for the eigenvalues and the
eigenstates [consider equations (\ref{v1}) -- (\ref{v3})], we see that for $%
\varepsilon _{\Omega }\ll \max \{g_{0},\Delta _{-}\}$ the
time-dependent eigenvalues and eigenstates are very close to the
time-independent ones evaluated at the bare qubit frequency
$\Omega _{0}$. In this work we assume a small modulation depth and
$g_{0}\ll \omega _{0},\Omega _{0}$, hence in the ME (\ref{dave})
we can use the lowest order {\em time-independent} Rabi eigenvalues and
eigenstates given by equations (\ref{v1}) -- (\ref{v3})~\footnote{In fact, the relative error we make in evaluating the Hamiltonian eigenstates under such hypotheses
is of the order of $\epsilon_\Omega/\max\{g_0,\Delta_-\}$ or smaller (from the expressions for
$\cos\theta_n$ and $\sin\theta_n$ it is easy to find out that the
relative error is of the order of $\epsilon_\Omega g_0/\Delta_-^2$
in the dispersive regime and $\epsilon_\Omega/g_0$ in the
resonant regime), and the relative error in the derivation
of the dissipator is of the same order.}. Moreover, we do not
restrict our analysis to a specific model for the reservoirs'
spectral densities and make the simplest assumption that the
dissipation rates are zero for $\varpi <0$ and take on constant values $%
\kappa $, $\gamma $ and $\gamma _{\phi }$ for $\varpi \geq 0$. Since our
primary goal is to study the photon generation from vacuum, such assumption
is the most conservative with regards to the spurious generation of
excitations due to dephasing \cite{blais,werlang,mendart}. Besides,
eventual random fluctuations in the modulation frequency may be treated as
additional dephasing noise \cite{blais,werlang}, so the simultaneous
inclusion of $\kappa $, $\gamma $ and $\gamma _{\phi }$ covers the most
common experimental situations.

The numeric integration of the master equation (\ref{dave}) with the
approximate Rabi eigenstates $|R_{i}\rangle $ is still cumbersome from the
practical viewpoint, so we also evaluate the kernel $\mathcal{\hat{L}}_{JC}$
in which one uses the time-independent Jaynes-Cummings eigenvalues and
eigenstates obtained by setting $\delta _{+}=\Lambda =\xi =0$ in equations (%
\ref{v1}) - (\ref{v3}). It will be shown that in our examples, where $%
\Lambda \approx 0.02$, these two approaches give almost indistinguishable
results, which are qualitatively similar to the prediction of the
phenomenological \lq standard master
equation\rq\ of Quantum Optics (in whose microscopic
derivation one assumes that the qubit and resonator do not interact \cite{vogel})
\begin{equation}
\mathcal{\hat{L}}_{ph}\bullet =\kappa \mathcal{D}[\hat{a}]\bullet +\gamma
\mathcal{D}[\hat{\sigma}_{-}]\bullet +\frac{\gamma _{\phi }}{2}\mathcal{D}[%
\hat{\sigma}_{z}]\bullet ~
\end{equation}%
with constant dissipative rates $\kappa $, $\gamma $ and $\gamma _{\phi }$.
So the knowledge of regimes in which $\mathcal{\hat{L}}_{ph}$ provides the
same results as $\mathcal{\hat{L}}_{R}$ may be important for future studies
where the numerical evaluation of $\mathcal{\hat{L}}_{R}$ or $\mathcal{\hat{L%
}}_{JC}$ is prohibitively complicated.

\section{Numerical results}

\label{sec3}

We verified the feasibility of photon generation from vacuum via
Landau-Zener sweeps of the modulation frequency in actual circuit QED setups
by solving numerically the master equation (\ref{mer}) for the kernels $%
\mathcal{\hat{L}}_{R}$, $\mathcal{\hat{L}}_{JC}$ and $\mathcal{\hat{L}}_{ph}$%
. We considered the standard value $\omega _{0}/2\pi =8\,$GHz for the cavity
frequency, the realistic qubit--field coupling strength $g_{0}/\omega
_{0}=4\times 10^{-2}$ and the currently available dissipative rates $\kappa
=10^{-4}g_{0}$, $\gamma =\gamma _{\phi }=7\times 10^{-4}g_{0}$ \cite{exp1,exp2,exp3}.

\begin{figure}[tbh]
\begin{center}
\includegraphics[width=0.4\textwidth]{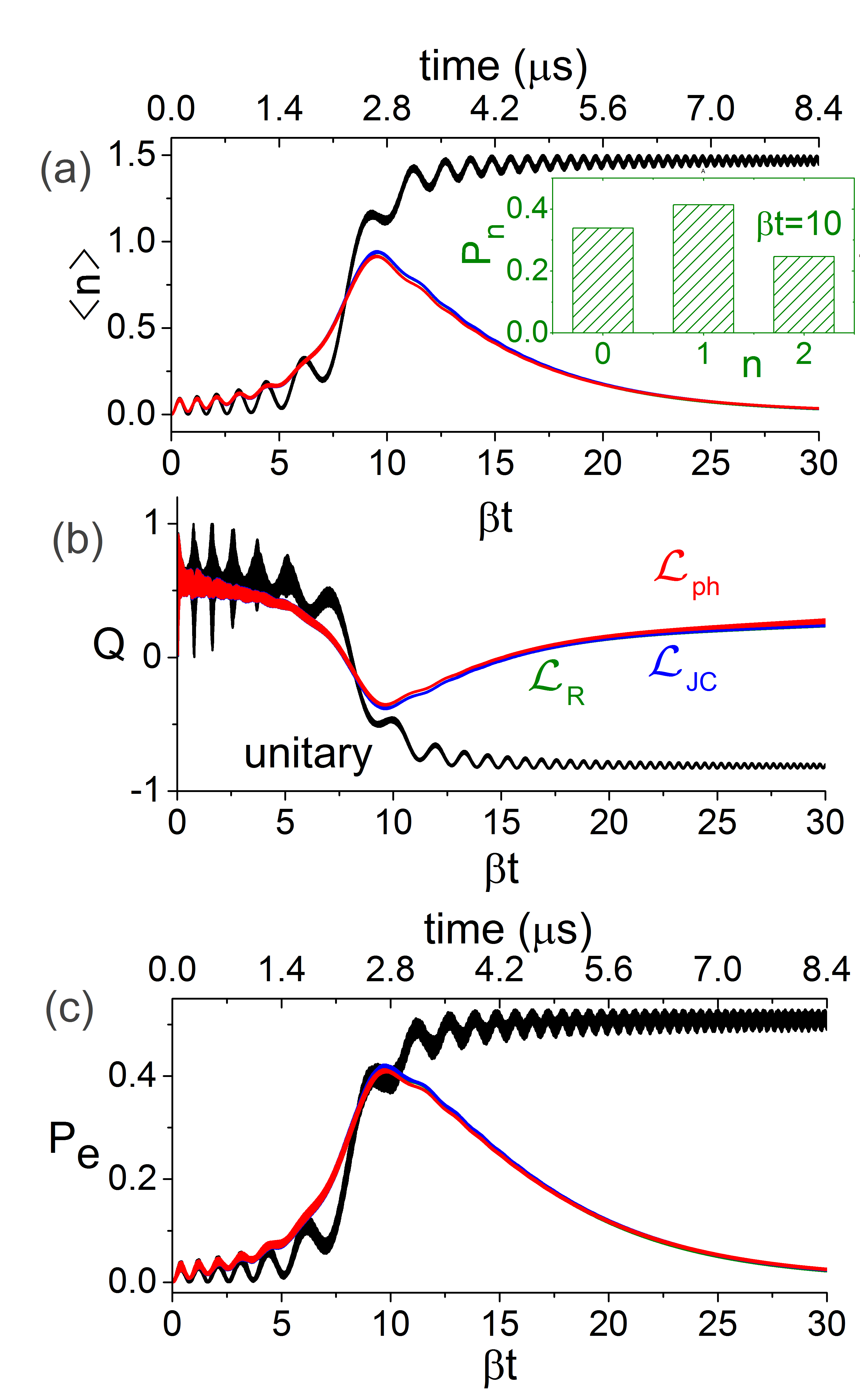} {}
\end{center}
\caption{(Color online) Time behavior of: a) average photon number $\langle \hat{n}\rangle $, b) Mandel's $Q$-factor and c) the atomic excitation probability $P_{e}$ in the
resonant regime for the modulation frequency $\protect\eta =2\protect\omega %
_{0}+g_{0}\protect\sqrt{2}-\protect\nu (t)$ and initial state $|g,0\rangle $%
. Black lines correspond to the unitary evolutions, while red, blue and green lines correspond to the three dissipative models characterized by ${\cal L}_{ph}$,${\cal L}_{JC}$,${\cal L}_{R}$, respectively. The three dissipative models give almost the same results.
The inset in (a) displays the photon statistics at the time instant $\protect%
\beta t=10$ under the dissipative kernel $\mathcal{L}_{R}$.}
\label{fig1}
\end{figure}

In figure \ref{fig1} we exemplify the implementation of the transition $%
|g,0\rangle \rightarrow |R_{2,+}\rangle $ in the resonant regime for the
parameters: $\Delta _{-}=0$, $\varepsilon _{\Omega }=0.01\times \Omega _{0}$%
, $\eta =2\omega _{0}+g_{0}\sqrt{2}-\nu (t)$, where $\nu (t)=-8\beta +(\beta
^{2}/2)t$ and $\beta \equiv g_{0}\varepsilon _{\Omega }/(2\sqrt{2}\Delta
_{+})$. We plot the average photon number $\langle \hat{n}\rangle $, the
Mandel's Q-factor $Q=[\langle (\Delta \hat{n})^{2}\rangle -\langle \hat{n}%
\rangle ]/\langle \hat{n}\rangle $ (that quantifies the spread of the photon
number distribution) and the atomic excitation probability $P_{e}=\mathrm{Tr}%
[|e\rangle \langle e|\hat{\rho}]$. In the ideal case the system ends up
approximately in the dressed state $|R_{2,+}\rangle $. Under realistic
conditions the photon generation from vacuum persists for initial times, but
later the system decays to the ground state associated to the kernel $%
\mathcal{\hat{L}}$ because the external modulation goes off-resonance. We
notice that the predictions of different dissipation models are quite
similar in this example, so for a estimative of the time behavior one can
employ the simplest phenomenological master equation. In the inset we show
the photon statistics evaluated at the time interval $\beta t=10$ for the
dissipator $\mathcal{L}_{R}$, which confirms that one or two photons could
be measured with roughly $70\,\%$ probability.

\begin{figure}[tbh]
\begin{center}
\includegraphics[width=0.4\textwidth]{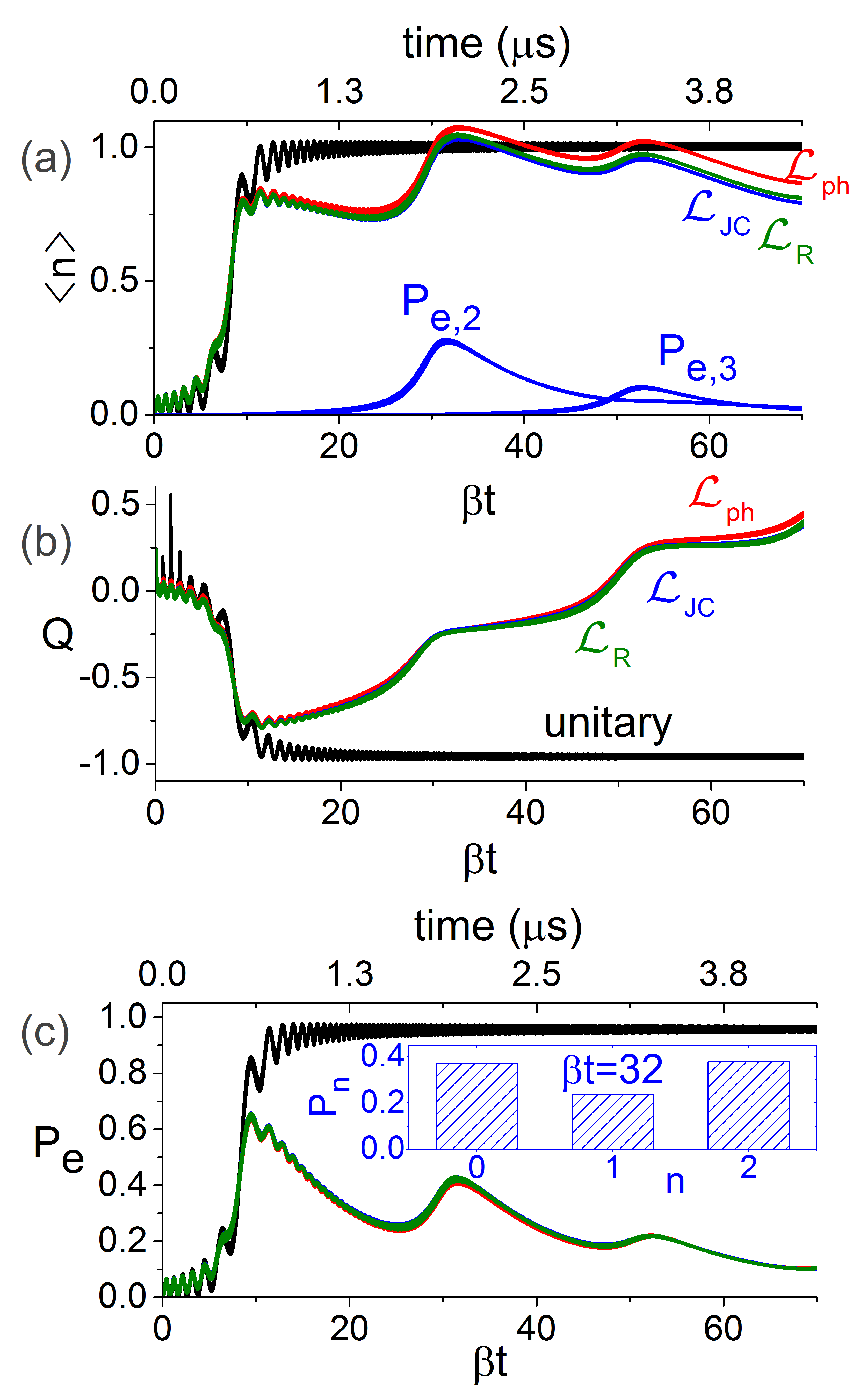} {}
\end{center}
\caption{(Color online) Behavior of: a) $\langle \hat{n}\rangle $, b) $Q$ and c) $%
P_{e}$ for the coupling between the states $|g,0\rangle \rightarrow |R_{2,-%
\mathcal{D}}\rangle $ in the dispersive regime and modulation frequency $%
\protect\eta =\Delta _{+}-2\left( \protect\delta _{-}-\protect\delta %
_{+}\right) +4\protect\alpha -\protect\nu (t)$. For the dissipator $\mathcal{%
L}_{JC}$ we also show the probabilities $P_{e,n}\equiv \mathrm{Tr}%
(|e,n\rangle \langle e,n|\hat{\protect\rho})$ (a) and the photon statistics for $%
\protect\beta t=32$ [inset in (c)]. }
\label{fig2}
\end{figure}

In figure \ref{fig2} we consider the transition $|g,0\rangle \rightarrow
|R_{2,-\mathcal{D}}\rangle $ in dispersive regime for the parameters: $%
\Delta _{-}=9g_{0}$, $\varepsilon _{\Omega }=0.04\times \Omega _{0}$, $\eta
=\Delta _{+}-2\left( \delta _{-}-\delta _{+}\right) +4\alpha -\nu (t)$,
where $\nu (t)=-8\beta +(\beta ^{2}/2)t$ and $\beta \equiv g_{0}\varepsilon
_{\Omega }/2\Delta _{+}$. Under the unitary evolution one would generate
approximately the state $|e,1\rangle $, but the dissipation alters
dramatically such behavior due to successive couplings $|R_{1,\mathcal{D}%
}\rangle \rightarrow |R_{3,-\mathcal{D}}\rangle $, $|R_{2,\mathcal{D}%
}\rangle \rightarrow |R_{4,-\mathcal{D}}\rangle $ (roughly $|g,1\rangle
\rightarrow |e,2\rangle $ and $|g,2\rangle \rightarrow |e,3\rangle $)
induced by the modulation for large times, while the transitions $|R_{2,-%
\mathcal{D}}\rangle \rightarrow |R_{1,\mathcal{D}}\rangle $ and $|R_{3,-%
\mathcal{D}}\rangle \rightarrow |R_{2,\mathcal{D}}\rangle $ are caused by
dissipation. These additional transitions are testified by plotting the
behavior of probabilities $P_{e,n}\equiv \mathrm{Tr}[|e,n\rangle \langle e,n|%
\hat{\rho}]$ obtained from the kernel $\mathcal{L}_{JC}$ (figure \ref{fig2}a); they can be
avoided by sweeping the modulation frequency in the opposite direction, $\nu
(t)\rightarrow -\nu (t)$, since in this case the transition $|R_{1,\mathcal{D%
}}\rangle \rightarrow |R_{3,-\mathcal{D}}\rangle $ becomes off-resonant for
larger times. In the inset of \ref{fig2}c we show the photon statistics obtained from the
kernel $\mathcal{L}_{JC}$ for the time instant $\beta t=32$, which proves
that two photons could be observed experimentally. Again we notice that the
predictions of the kernels $\mathcal{L}_{JC}$ and $\mathcal{L}_{R}$ are
almost indistinguishable, and qualitatively identical to the predictions of $%
\mathcal{L}_{ph}$. So in the remaining of the paper we shall only employ the
dissipators $\mathcal{\hat{L}}_{JC}$ and $\mathcal{\hat{L}}_{ph}$, as the
evaluation of $\mathcal{\hat{L}}_{R}$ becomes too demanding for a large
number of excitations.

\begin{figure}[tbh]
\begin{center}
\includegraphics[width=0.4\textwidth]{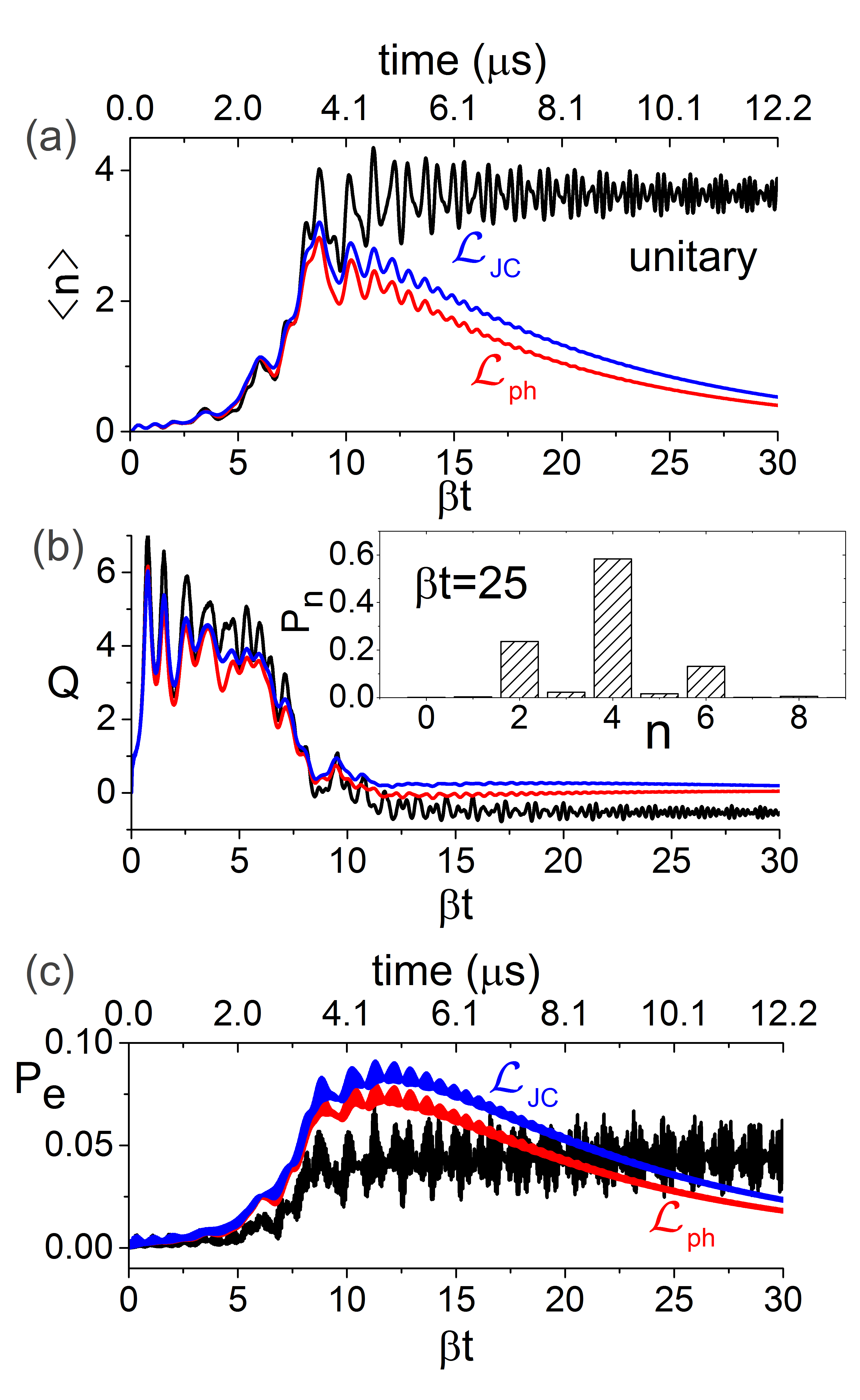} {}
\end{center}
\caption{(Color online) Coupling of the states $|g,0\rangle \rightarrow
|R_{2k,\mathcal{D}}\rangle $ in the dispersive regime for the modulation
frequency $\protect\eta =2\protect\omega _{0}+2\left( \protect\delta _{-}-%
\protect\delta _{+}\right) -4\protect\alpha -\protect\nu (t)$. Inset in (b): photon
statistics under the unitary evolution for the time instant $\protect\beta %
t=25$.}
\label{fig3}
\end{figure}
\begin{figure}[tbh]
\begin{center}
\includegraphics[width=0.4\textwidth]{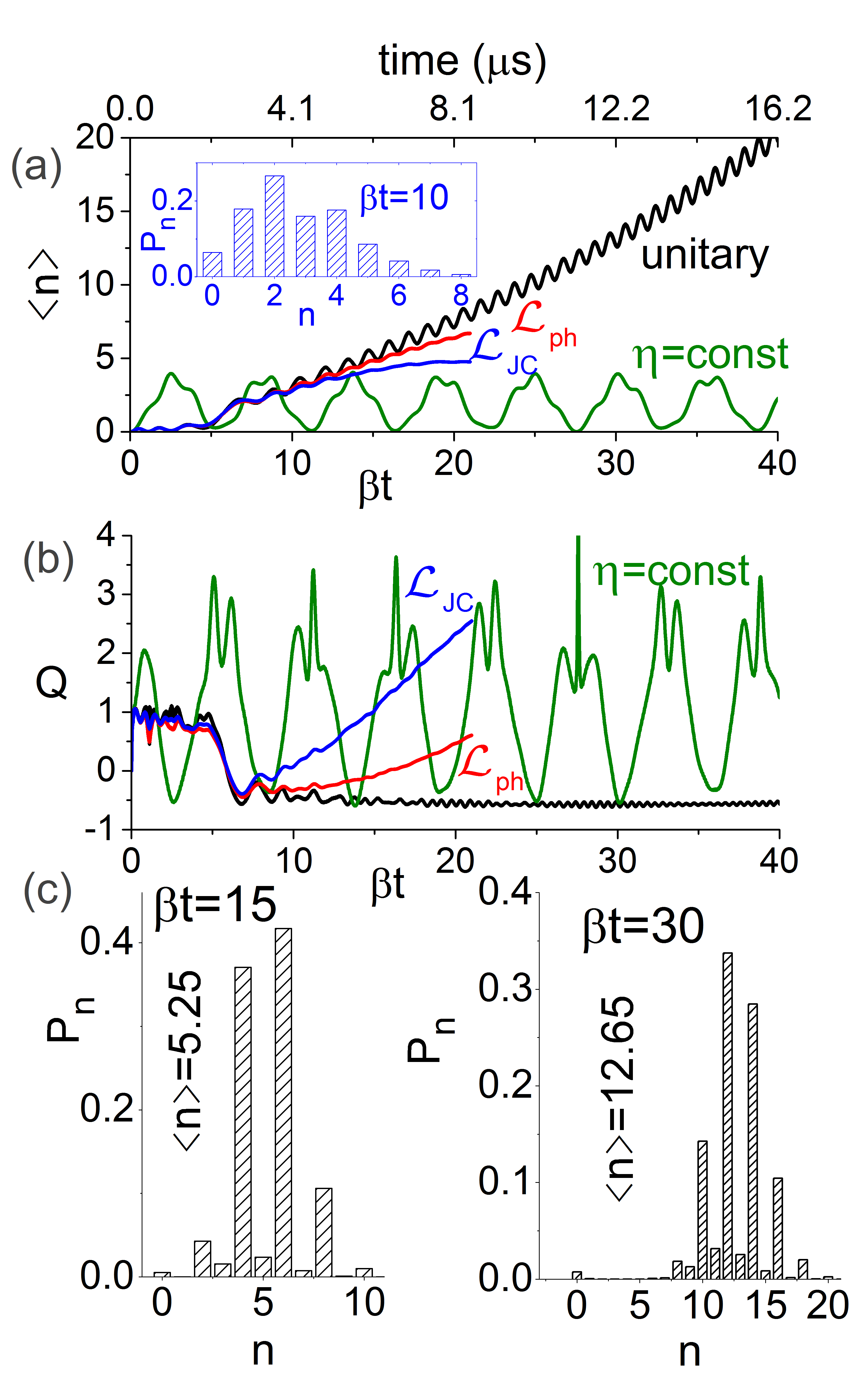} {}
\end{center}
\caption{(Color online) Similar to figure \protect\ref{fig3} but for the
modulation frequency $\protect\eta =2\protect\omega _{0}+2\left( \protect%
\delta _{-}-\protect\delta _{+}\right) -4\protect\alpha +\protect%
\nu (t)$. a) and b): time behavior of $\langle \hat{n} \rangle$ and $Q$ under different dissipators, where the green lines labeled \lq $\protect\eta =const$%
\rq\ correspond to harmonic modulation with constant
frequency adjusted to maximize $\langle \hat{n}\rangle $. Inset in (a): photon
statistics for $\protect\beta t=10$ under dissipator $\mathcal{L}_{JC}$.
c) photon statistics under unitary evolution for two time
instants: $\protect\beta t=15$ and $30$.}
\label{fig4}
\end{figure}

In figures \ref{fig3} -- \ref{fig4} we consider multiple transitions $%
|g,0\rangle \rightarrow |R_{2k,\mathcal{D}}\rangle $ ($k=1,2,\ldots $) in
dispersive regime for the parameters of figure \ref{fig2} and the modulation
frequency $\eta =2\omega _{0}+2\left( \delta _{-}-\delta _{+}\right)
-4\alpha -\mathcal{S}\nu (t)$, where $\nu (t)=[8\beta -(\beta ^{2}/2)t]$, $%
\mathcal{S}=\pm $ and $\beta \equiv \delta _{-}\varepsilon _{\Omega }/\sqrt{2%
}\Delta _{+}$. For these parameters we have $\beta /\alpha \approx 0.9$, so
from the Hamiltonian (\ref{mart}) we expect Landau-Zener transitions among
several dressed states during each avoided-crossing. In figure \ref{fig3} we
set $\mathcal{S}=+$, so only the states $|R_{2k,\mathcal{D}}\rangle $ with $%
k\sim 1$ may become populated from the initial vacuum state, as can be seen
from the plots. In the ideal case the atom acquires a small probability of
excitation and up to six photons are created with non-negligible probability
-- this is shown in the inset of \ref{fig3}b, where the photon statistics is displayed for
the time instant $\beta t=25$. Besides, the created field state is
nonclassical and quite different from the usual squeezed vacuum state
created during standard DCE, as corroborated by the negative values of the $%
Q $-factor (recall that for the SVS one has $Q=1+2\langle \hat{n}\rangle $).
In the presence of dissipation the system tends to the ground state for
large times, since the modulation becomes off-resonant. Still it is possible
to measure a meaningful average photon number $\langle \hat{n}\rangle \sim
2 $ on the timescales of a few microseconds.

When $\mathcal{S}=-$ one can generate
a quite large amount of photons from vacuum in the ideal case, since the system undergoes
successive Landau-Zener transitions towards the higher-energy dressed
states. This is illustrated in figure \ref{fig4}. For adiabatic variation of $\eta$ only a few photon states are populated at a time (see \ref{fig4}c for the photon statistics at the time instants $%
\beta t=15$ and $30$), which is reflected in the sub-Poissonian photon
statistics for large times. The amount of created photons can substantially
exceed the average photon number achievable for the harmonic modulation with
constant frequency \cite{igor,diego} (green lines in \ref{fig4}a and \ref{fig4}b, for which $\eta $ was
found numerically to optimize the photon generation). Our numerical
simulation of dissipation is not accurate for large photon numbers, so we
only show the dissipative dynamics for $\beta t<21$. The differences
between the predictions of kernels $\mathcal{L}_{JC}$ and $\mathcal{L}_{ph}$
become significant for $\langle \hat{n}\rangle \gtrsim 4$ due
to the differences in the available decay channels, but the overall behavior
is qualitatively similar. Our results demonstrate that several photons can
be generated on the timescales of a few microseconds, but the dissipation
modifies the photon statistics to super-Poissonian (see the inset in \ref{fig4}a, evaluated at
the time instant $\beta t=10$ for the kernel $\mathcal{L}_{JC}$).

\begin{figure}[tbh]
\begin{center}
\includegraphics[width=0.48\textwidth]{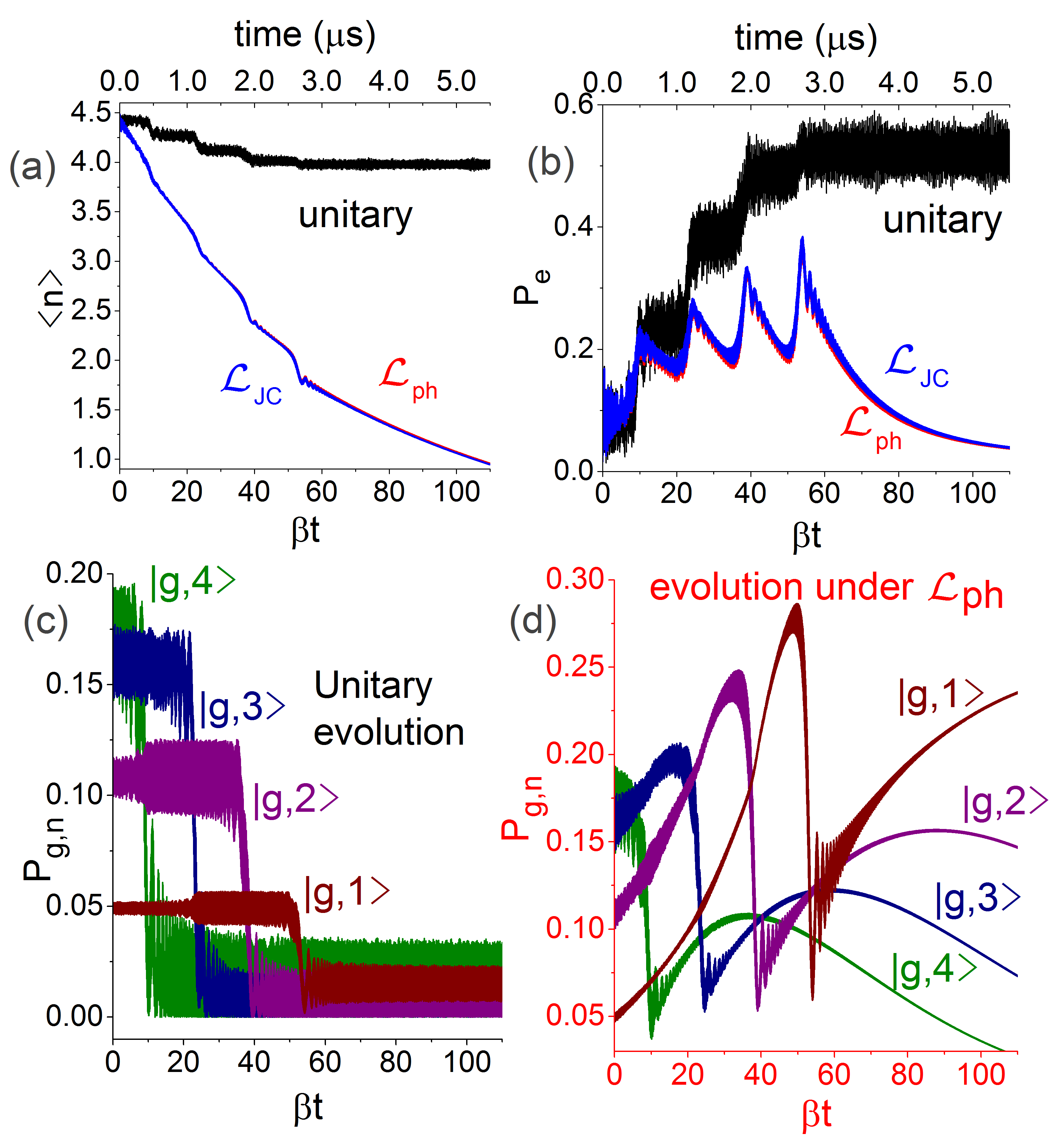} {}
\end{center}
\caption{(Color online) Coupling of the states $|R_{n,+}\rangle \rightarrow
|R_{n,-}\rangle $ for $n\leq 4$ in the dispersive regime for the initial
state $|g,\protect\alpha \rangle $ and modulation frequency $\protect\eta %
=\Delta _{-}+2M\left( \protect\delta _{-}-\protect\delta _{+}\right) -%
\protect\nu (t)$. a) and b): time behavior of $\langle \hat{n}\rangle$ and $P_e$ under different dissipators. c) and d): behavior of probabilities $P_{g,n}\equiv \mathrm{Tr%
}(|g,n\rangle \langle g,n|\hat{\protect\rho})$ under the unitary evolution and
the kernel $\mathcal{L}_{ph}$, respectively.}
\label{fig5}
\end{figure}

Finally, in figure \ref{fig5} we study the transitions between the dressed
states $|R_{n,+}\rangle \rightarrow |R_{n,-}\rangle $ in the dispersive
regime for $n\leq M$, which roughly correspond to the transitions $%
|g,n\rangle \rightarrow |e,n-1\rangle $. We consider the initial state $%
|g,\alpha \rangle $, where $|\alpha \rangle $ denotes the coherent state
with $\alpha =\sqrt{4.5}$. Other parameters are: $\Delta _{-}=10g_{0}$, $%
\varepsilon _{\Omega }/\omega _{0}=4\times 10^{-3}$, $\eta =\Delta
_{-}+2M\left( \delta _{-}-\delta _{+}\right) -\nu (t)$, where $\nu
(t)=-7\beta +(\beta ^{2}/2)t$, $M=4$ and $\beta \equiv g_{0}\varepsilon
_{\Omega }/\Delta _{-}$. As expected, the dissipation strongly affects the
dynamics, since excitations are lost to the environment from the very
beginning and the slow modulation is unable to create additional
excitations. To confirm the selective Landau-Zener sweeps between the states
$\{|R_{n,+}\rangle ,|R_{n,-}\rangle \}$ in \ref{fig5}c and \ref{fig5}d we show the time evolution of the
probabilities $P_{g,n}\equiv \mathrm{Tr}(|g,n\rangle \langle g,n|\hat{\rho})$%
, which change one at a time from the initial value to roughly zero ($%
P_{g,n} $ also undergoes fast oscillations due to the dispersive exchange of
excitations between the field and the qubit). For $n>M$ the probabilities $%
P_{g,n}$ are not affected by the perturbation, since the corresponding
avoided-crossings are not swept during the frequency variation (for the
frequency change in the opposite directions, $\nu (t)\rightarrow -\nu (t)$,
only the states with $n\geq M$ would be affected). Although under
dissipation the occurrence of the Landau-Zener transitions is almost completely
washed out in the behavior of $\langle \hat{n}\rangle $ (figure \ref{fig5}a), the atomic
excitation probability $P_{e}$ still preserves the characteristic
Landau-Zener plateaus, which are transformed into peaks due to the damping (figure \ref{fig5}b).

\section{Discussion and conclusions}

\label{sec4}

In this work we  have discussed a simple scheme to achieve photon generation from
vacuum due to the counter-rotating terms in a time-dependent
Rabi Hamiltonian, where a suitable perturbation characterized by a sinusoidal time dependence with a slowly changing frequency is present.
 Because of this frequency change the perturbation intercepts several resonance frequencies of the system (described by the time-independent Rabi Hamiltonian) and can induce a single or a series of Landau-Zener processes that allow for populating the upper levels starting from the ground state.
This scheme then works quite well without the need to know accurately the resonant unperturbed frequencies of the system nor accurately adjusting the shape of the modulation frequency. Besides, it is only
little sensitive to the duration of the external perturbation. The latter property is related to the fact that, provided the diabatic energies (i.e., the diagonal Rabi-basis matrix elements) in the effective Hamiltonians vary in a sufficiently wide range and with a sufficiently low speed, each Landau-Zener process essentially leads to the same final result, irrespectively of the specific range and speed.

It is worth noting that although we considered the
modulation of the atomic transition frequency, our method can
easily be extended to the time-variation of the atom-field coupling
strength, or both. Moreover, the effective Hamiltonians deduced here are valid for
arbitrary variation of the modulation frequency. So our results can be
employed to propose protocols that optimize, for instance, the photon production or
generation of specific entangled states, although in these cases one would
need a precise knowledge of the system spectrum and to control accurately
the duration of the perturbation.

We showed that the effective Landau-Zener transitions could be implemented
for current parameters of dissipation provided one maintains the modulation
for a time interval of the order of microseconds with relative modulation
strength of a few percent. For consistent description of the experimental
results the dissipation must be necessarily included because it alters
significantly the unitary behavior due to the additional transitions between
the system eigenstates induced by the combined action of dissipation and
coherent perturbation. It is remarkable that the predictions of the
phenomenological quantum optical master equation are qualitatively similar
to the predictions of the microscopic dressed-picture master equation, and
in many situations both predictions are almost indistinguishable (this is typical in the weak damping limit). This means that one can use the simpler phenomenological master equation to get a
crude estimation about the overall behavior.  The effect of random
fluctuations of the modulation frequency can be incorporated into our approach
as additional pure dephasing, so one does not expect major qualitative differences in
this case. Therefore we hope that our protocol will facilitate the hitherto missing experimental observation of DCE due to a single qubit.

\begin{acknowledgments}
A.V.D. would like to thank Dipartimento di Fisica e Chimica of  Universit\`a degli studi di Palermo for the warm
hospitality during the visit in winter of 2016. A.V.D. also acknowledges a support of Brazilian agencies CNPq (Conselho Nacional
de Desenvolvimento Cient\'{\i}fico e Tecnol\'{o}gico) and FAPDF (Funda\c{c}%
\~{a}o de Apoio a Pesquisa do Distrito Federal). \end{acknowledgments}

\end{document}